# Floquet engineering of quantum relaxation

**Kang Xu[1], Dionisio Cendoya[3,4], Nazar Delegan[5,6], F. Joseph Heremans[5,6], Olga Shenderova[7], Daniela Pagliero[1,2], Lisandro Buljubasich[3,4,†], and Carlos A. Meriles[1,2,†]**

[1]*Department of Physics, CUNY- The City College of New York, New York, NY 10031, USA.*
[2]*CUNY-Graduate Center, New York, NY 10016, USA.*
[3]*Facultad de Matemática, Astronomía, Física y Computación, Universidad Nacional de Córdoba, Córdoba, Argentina.*
[4]*CONICET, Instituto de Física Enrique Gaviola (IFEG), Córdoba, Argentina.*
[5]*Materials Science Division and Q-NEXT, Argonne National Laboratory, Lemont, IL 60439, USA.*
[6]*Pritzker School of Molecular Engineering, University of Chicago, Chicago, IL 60637, USA.*
[7]*Adamas Nanotechnologies, Inc., Raleigh, NC 27617, USA.*

[†]E-mail: lisandro.buljubasich@unc.edu.ar, cmeriles@ccny.cuny.edu

Dynamical decoupling suppresses decoherence by averaging selected interactions between a quantum system and its environment. Here, we extend this notion through a general Floquet formalism that maps periodic control onto relaxation and dephasing channels. Environmental fluctuations are decomposed into Floquet components that sample the noise spectrum at distinct frequencies, with coupling strengths determined by control-induced Floquet coefficients. We experimentally investigate this approach using a single nitrogen-vacancy center in diamond under multi-pulse control. By tuning the drive and detuning, we reshape the coupling to environmental fluctuations and create isolated operating points where the dominant decoherence channel vanishes, strongly suppressing Floquet dephasing. The resulting coherence enhancement follows the predicted Floquet coupling coefficients and can be shifted by modifying the control sequence. These results establish Floquet driving as a general framework for engineering transverse and longitudinal relaxation through the design of quantum control protocols.

Decoherence arises from the coupling of a quantum system to fluctuations in its environment and constitutes a central limitation in quantum information processing, sensing, and spectroscopy. A powerful strategy to mitigate its effects is dynamical decoupling, where appropriately timed control operations average selected system-environment interactions while preserving the desired quantum evolution[1-3]. Closely related approaches, including continuous driving and spin locking[4-8], can additionally shift the system response away from low-frequency noise, providing both enhanced coherence and access to environmental spectral information. These methods have become essential tools for protecting quantum states and characterizing their surroundings[9-11].

A driven quantum system is best described in terms of Floquet states, whose coupling to environmental fluctuations is generally redistributed among multiple harmonics and between operators that either preserve or change the Floquet-state populations[12-15]. The resulting relaxation and dephasing processes therefore depend not only on the environmental noise spectrum, but also on the structure imposed by the periodic control[16]. This suggests a broader viewpoint in which the system-environment interaction is reshaped to determine how different portions of the environmental spectrum contribute to the dynamics[9].

Establishing such control requires a framework connecting an arbitrary periodic drive to the corresponding dissipative dynamics. While Floquet theory provides a natural description of periodically driven systems, its use in open quantum systems has largely focused on predicting relaxation, heating, or steady-state behavior for a prescribed drive[17-19]. For quantum control, the inverse perspective is particularly useful: Identifying how the control protocol determines the coupling coefficients and spectral frequencies governing relaxation and dephasing. Such a connection allows periodic driving to serve as a means of engineering decoherence channels, thus complementing existing coherence protection strategies.

Here, we develop a general Floquet formalism that establishes this connection and experimentally explore it using a single nitrogen-vacancy center in diamond under multi-pulse control. We express environmental interactions in the Floquet basis as a set of control-dependent coupling coefficients that separate population-relaxation and dephasing channels and determine the frequencies at which each samples the environmental spectrum. Using multi-pulse control, we continuously modify these coefficients through the drive parameters and identify isolated operating points at which the dominant dephasing coupling vanishes. Under these conditions, the Floquet coherence is strongly enhanced at selective locations defined by the chosen control protocol and operating parameters.

*Floquet engineering of system–environment coupling.* Consider a two-level system subject to periodic control and coupled weakly to a fluctuating environment,

$$\mathcal{H}(t) = \mathcal{H}_{\rm s} + \mathcal{H}_{\rm rf}(t) + \mathcal{H}_{\rm int}(t)\,. \qquad (1)$$

Here, $\mathcal{H}_{\rm s} = \omega_0 I_0$ is the bare system Hamiltonian with $I_0 \equiv I_z$ denoting the global spin operator along the main quantization axis, $\mathcal{H}_{\rm rf}(t)$ is the periodic control field with carrier frequency $\omega_{\rm rf}$ and cycle time $t_{\rm c}$, and $\mathcal{H}_{\rm int}(t)$ describes environmental fluctuations. In the rotating frame, the control generates the periodic transformation

$$\mathcal{P}(t) = \mathcal{T}\exp\left(-i\int_0^t dt'\mathcal{H}^*_{\rm rf}(t')\right), \qquad (2)$$

where $\mathcal{T}$ is the time-ordering operator, and the asterisk denotes a rotating frame transformation. The evolution generated by the detuning $\omega = \omega_0 - \omega_{\rm rf}$ in this toggling frame admits the Floquet representation[20]

$$\mathcal{Q}(t) = \mathcal{T}\exp\left(-i\int_0^t dt'\,\omega(t)\tilde{I}_z(t')\right) = \mathcal{F}(t)e^{-i\mathcal{H}_{\rm F}t}, \qquad (3)$$

where $\tilde{I}_z(t') = \mathcal{P}^\dagger(t)I_z\mathcal{P}(t)$, $\mathcal{F}(t)$ denotes a periodic operator chosen so that $\mathcal{F}(kt_{\rm c}) = 1$ at multiples $k = \{0,1,\ldots\}$ of the cycle time $t_{\rm c}$, and $\mathcal{H}_{\rm F} = \mathbf{\Omega}_{\rm F}\cdot\mathbf{I}$ is the Floquet Hamiltonian.

Environmental fluctuations couple to the system through spin operators $I_\alpha$, where $\alpha = \{0,+,-\}$ labels the longitudinal and transverse components. Under periodic control, each operator acquires a time-dependent representation that can be expanded in harmonics of the cycle frequency $\omega_{\rm c} = 2\pi/t_{\rm c}$,

$$\mathcal{F}^\dagger(t)\mathcal{P}^\dagger(t)I_\alpha\mathcal{P}(t)\mathcal{F}(t) = \sum_m e^{im\omega_{\rm c}t}I_\alpha^{(m)}\,, \qquad (4)$$

where

$$I_\alpha^{(m)} = \frac{1}{t_{\rm c}}\int_0^{t_{\rm c}} dt\,e^{-im\omega_{\rm c}t}\,\mathcal{F}^\dagger(t)\mathcal{P}^\dagger(t)I_\alpha\mathcal{P}(t)\mathcal{F}(t) = \sum_\beta c_{\alpha,\beta}^{(m)}J_\beta\,. \qquad (5)$$

In the last expression, $J_\beta$ with $\beta = \{0,+,-\}$ denotes the set of Floquet basis operators; the coefficients $c_{\alpha,\beta}^{(m)} = \mathrm{Tr}\left\{J_\beta^\dagger I_\alpha^{(m)}\right\}/\mathrm{Tr}\left\{J_\beta^\dagger J_\beta\right\}$ specify the amplitude with which the original coupling operator $I_\alpha$ is mapped onto Floquet operator $J_\beta$ at harmonic $m$. Crucially, because $\mathcal{P}(t)$, $\mathcal{F}(t)$, and the Floquet basis are set by the applied control, the coefficients $c_{\alpha,\beta}^{(m)}$ are themselves control-dependent quantities. In the interaction picture of $\mathcal{H}_{\rm F}$, each Floquet operator acquires an additional phase associated with the quasienergy splitting. Including the rotating-frame transformation of the original operator $I_\alpha$, the environmental coupling contains components oscillating at $\Omega_{\alpha,\beta}^{(m)} = \alpha\omega_{\rm rf} + m\omega_{\rm c} + \beta\Omega_{\rm F}$. Thus, each coefficient $c_{\alpha,\beta}^{(m)}$ defines both the strength of a coupling channel and the frequency at which that channel samples the environmental noise spectrum.

Using $\lambda(t)$ to denote a stochastic process with zero mean, we now consider the dominant transition-frequency fluctuations, here described by $\mathcal{H}_{\rm int}(t) = \lambda_0(t)I_0$, and define their two-sided spectral density as $\Lambda_0(\omega) = \int_{-\infty}^{\infty} d\tau\,e^{i\omega\tau}\langle\lambda_0(\tau)\lambda_0(0)\rangle$. Periodic control redistributes this single environmental coupling among the Floquet operators $J_\beta$. Components proportional to $J_0$ preserve the Floquet-state populations and produce pure dephasing, whereas those proportional to $J_\pm$ induce population relaxation. Using $\delta$ to highlight the "dynamic" nature of the protocol (as opposed to the more common "rotating frame" representation denoted with $\rho$), the corresponding relaxation rates can be written as[20]

$$\Gamma_{\phi\delta} = \frac{1}{2}\sum_m \left|c_{0,0}^{(m)}\right|^2\Lambda_0\left(\Omega_{0,0}^{(m)}\right), \qquad (6)$$

and

$$\Gamma_{1\delta} = \sum_{m,\beta=\pm}\left|c_{0,\beta}^{(m)}\right|^2\Lambda_0\left(\Omega_{0,\beta}^{(m)}\right), \qquad (7)$$

where we assumed the limit where the open system admits a time-independent Floquet Lindbladian[24]. From the above equations, it follows that quantum coherence under periodic control is governed by the decay rate

$$\Gamma_{2\delta} = \frac{\Gamma_{1\delta}}{2} + \Gamma_{\phi\delta}\,, \qquad (8)$$

where we defined $\Gamma_{i\delta} \equiv 1/T_{i\delta}$, $i = 1,2$ as the inverse longitudinal and transverse relaxation times, respectively. Equations (6) through (8) generalize the conventional description of relaxation and dephasing in driven quantum systems to arbitrary periodic control.

*Experimental study with an NV center.* We investigate these ideas using a 25 nm deep nitrogen-vacancy (NV) center in diamond[25] featuring a Ramsey coherence lifetime $T_2^* = 2.7$ μs[20]. Unlike conventional dynamical-decoupling protocols—designed to refocus the detuning accumulated during the time separating pulses—a Floquet sequence may use non-commuting rotations to generate a nontrivial, detuning-dependent evolution. The particular control protocol considered herein—referred to as DSL-4—consists of repeated $\pi/2$ rotations about different axes in the transverse plane, separated by free-evolution intervals $\tau$ (Fig. 1a). The sequence derives from the WaHuHa family of multipulse protocols originally developed for dipolar decoupling[26,27], and is not expected to extend the spin coherence lifetime for the individually-addressable, non-interacting system of NVs studied herein.

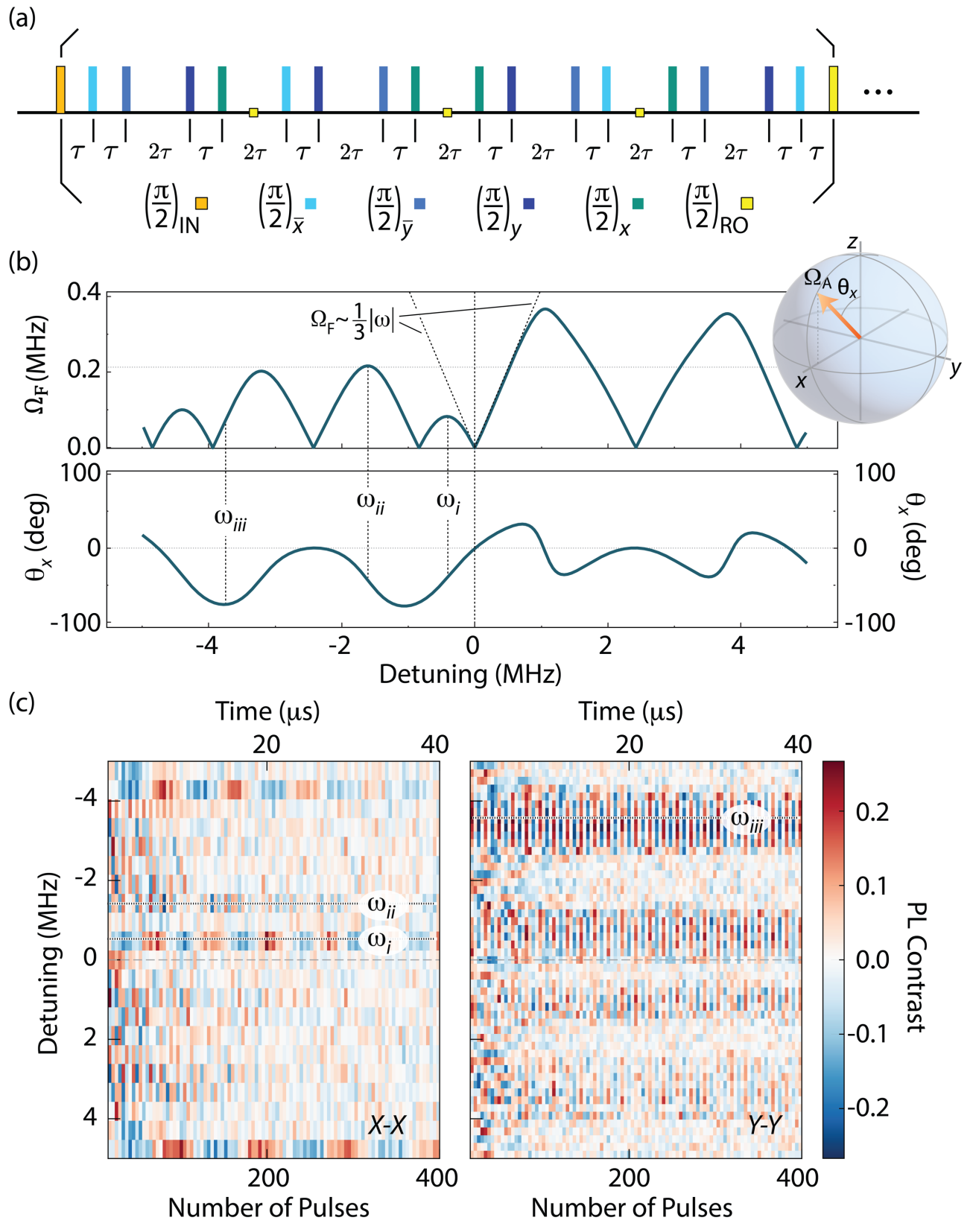

**FIG. 1: Floquet dynamics under periodic multipulse control.** (a) MW control sequence comprising alternating $\pi/2$ rotations separated by free-evolution intervals. Initialization (readout) MW pulses of variable phase are preceded (followed) by initialization (readout) optical pulses (not shown); yellow squares indicate optional readout points used to sample the micromotion. (b) Calculated Floquet-field amplitude $\Omega_\mathrm{F}$ and orientation $\theta_x$ versus detuning. Near resonance, $\Omega_\mathrm{F} \approx \omega/3$ (dashed line); for all detunings, the field orientation lies within the *xz* plane. (c) PL contrast versus detuning and pulse number, including all readout points in (a). In (b) and (c), we set $\tau = 100$ ns. IN: Initialization; RO: Readout.

In the small-detuning limit, dynamic averaging transforms the static offset into an effective field along $z$, with $\mathcal{H}_\mathrm{F} \approx \omega I_0/3$. Away from this limit, however, the Floquet field, which remains confined to the *xz* plane[20], develops a strongly nonlinear amplitude and orientation dependence (Fig. 1b), reflecting the breakdown of the average-Hamiltonian description.

Figure 1c shows the corresponding experimental dynamics, optionally sampled at intermediate points within the cycle (yellow markers in Fig. 1(a)) to resolve the micromotion. *X*-*X* measurements—i.e., experiments where the initialization and readout π/2-pulses have phases along *x*—predominantly probe coherence perpendicular to the Floquet axis, whereas *Y*-*Y* also detects polarization along it, realizing dynamic spin locking[28] when the two are aligned.

The pronounced detuning dependence of both signals can be qualitatively understood from the Floquet field in Fig. 1b. Detuning fluctuations perturb both its magnitude and orientation: fluctuations in $\Omega_\mathrm{F}$ produce quasienergy fluctuations and hence dephasing, whereas rotations of the Floquet axis induce population relaxation. Consequently, the distinct responses near $\omega_i$, and $\omega_{iii}$ reflect suppression of different decoherence channels: the long-lived *X*-*X* coherence near $\omega_i$ is associated with reduced Floquet dephasing, while the persistent *Y*-*Y* polarization near $\omega_{iii}$ signals suppressed population relaxation. Both persist far beyond the unprotected dynamics and motivate the quantitative Floquet analysis below.

To better identify the conditions required for long-lived coherent dynamics, we examine the signal's Fourier spectra over progressively longer acquisition windows, as illustrated via the *Y*-*Y* dynamics in Fig. 2a. For sequences with the shorter number of pulses $N$, the spectrum contains multiple branches reflecting the quasienergy evolution. The vertical features at $1/t_\mathrm{c}$ arise from spin-locked polarization, whereas the branches emerging near zero frequency follow the approximately linear Floquet response expected at small detuning, to subsequently break into the nonlinear structure anticipated from Fig. 1b. Increasing $N$ (relatively) suppresses short-lived spectral components, such that the $N = 400$ spectrum retains only a discrete set of narrow features at selected detunings. The spectroscopy thus reveals directly that the coherence lifetime is itself strongly detuning dependent. For example, Fig. 2b highlights one such long-lived feature at $\omega_{ii}$, probed selectively over a narrower detuning range using *X*-*X* initialization and readout. The time-domain data show a sharply localized region of extended coherence, while the corresponding spectrum resolves the associated Floquet branch.

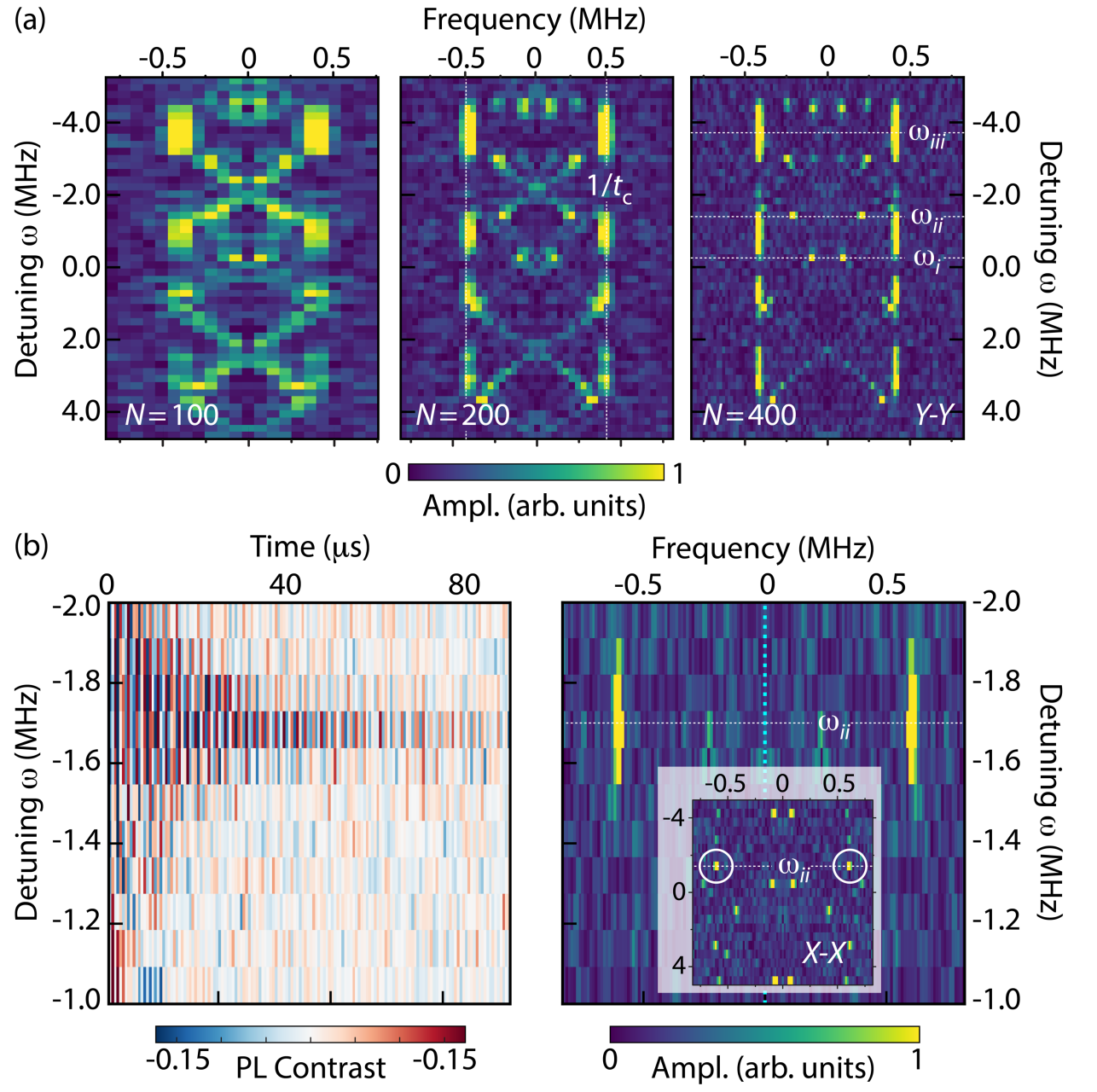

**FIG. 2: Detuning-selective Floquet coherence.** (a) Fourier spectra after a variable number of pulses $N$. Vertical features at $1/t_\mathrm{c}$ correspond to spin-locked polarization, while the low-frequency branches track the detuning-dependent Floquet evolution. Increasing $N$ and isolates narrow, long-lived features at selected detunings, including $\omega_i$ through $\omega_{iii}$. (b) Expanded *X*-*X* measurement around$\omega_{ii}$: time-domain PL contrast (left) and corresponding Fourier spectrum (right). The inset shows the full X-X spectrum, with circles marking the long-lived regions selected for detailed study.

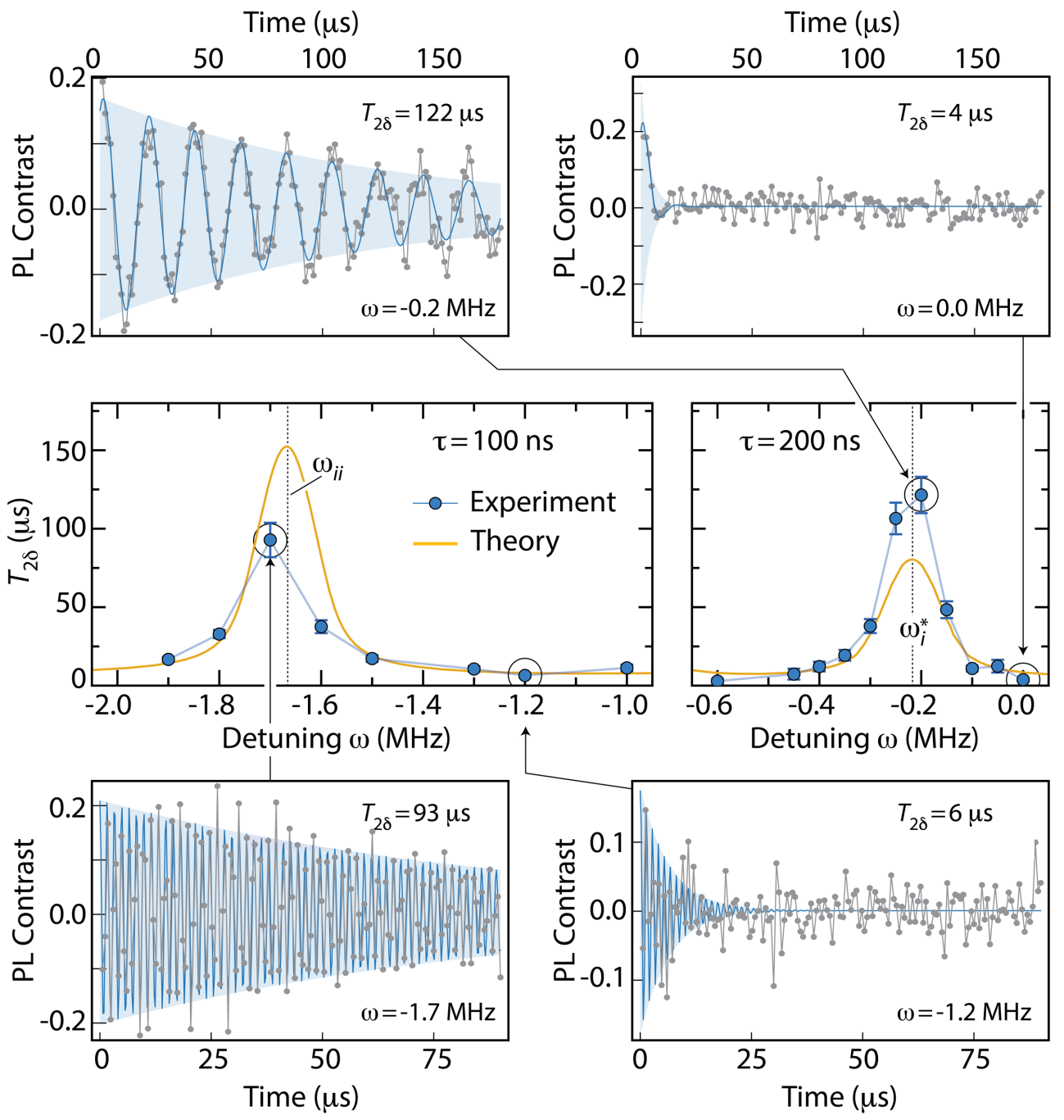

**FIG. 3: Floquet engineering of decoherence.** (Center) Measured $T_{2\delta}$ versus detuning around two protected regions identified in Fig. 1. Orange traces are predictions from Eqs. (6)–(8), calculated using the independently measured noise spectrum $S_0(\omega)$ and Floquet coupling coefficients $c_{0,\beta}^{(m)}$. Representative *X*-*X* traces at the coherence maxima show $T_{2\delta} = 122$ μs (top) and 93 μs (bottom); nearby detunings yield $T_{2\delta} \sim 4 - 6$ μs. The asterisk in $\omega_i^*$ indicates the scaling expected from switching $\tau$ from 100 to 200 ns.

*Engineering Floquet relaxation.* We next turn to a quantitative description of these long-lived regions using the Floquet framework developed above. For transition-frequency fluctuations, the coefficient governing the $m = 0$ Floquet dephasing channel has a particularly simple interpretation. Applying the Floquet–Hellmann–Feynman relation to the detuning-dependent quasienergies[20] gives $c_{0,0}^{(0)} = \partial\Omega_{\mathrm{F}}/\partial\omega$. Thus, stationary points of the quasienergy splitting correspond directly to zeros of the dominant low-frequency dephasing coupling. Importantly, this does not imply complete decoupling from the environment: Finite-$m$ dephasing terms $c_{0,0}^{(m)}$, as well as the population-relaxation terms $c_{0,\pm}^{(m)}$, are generally non-null and sample the noise spectrum at their corresponding Floquet frequencies.

Figure 3 tests this description quantitatively. We independently measure the transition-frequency noise spectrum $S_0^{(\mathrm{NV})}(\omega)$ of the NV center[20] and, using the calculated coefficients $c_{0,\beta}^{(m)}$, evaluate Eqs. (6)–(8) to predict $T_{2\delta}$ as a function of detuning. The calculation reproduces reasonably well both the position and width of the narrow coherence windows, as well as the absolute enhancement of $T_{2\delta}$, which reaches 122 and 93 μs in the two regions shown. Only a few hundred kilohertz away, the $m = 0$ dephasing is restored and $T_{2\delta}$ falls to ~4 μs, as illustrated by the representative traces.

Since both the coupling coefficients and the frequencies at which they sample the environment are determined by the periodic control, the individual contributions to relaxation and decoherence can themselves be engineered. Figure 4 numerically illustrates this capability for a model environment with Lorentzian spectral density $S_0^{(\mathrm{L})}(\omega)$. For the sequence used above, Fig. 4a decomposes the calculated population-relaxation and dephasing rates into their individual Floquet coupling coefficients. Note that suppression of the $m = 0$ dephasing channel need not suppress population relaxation, nor does it eliminate dephasing through higher harmonics.

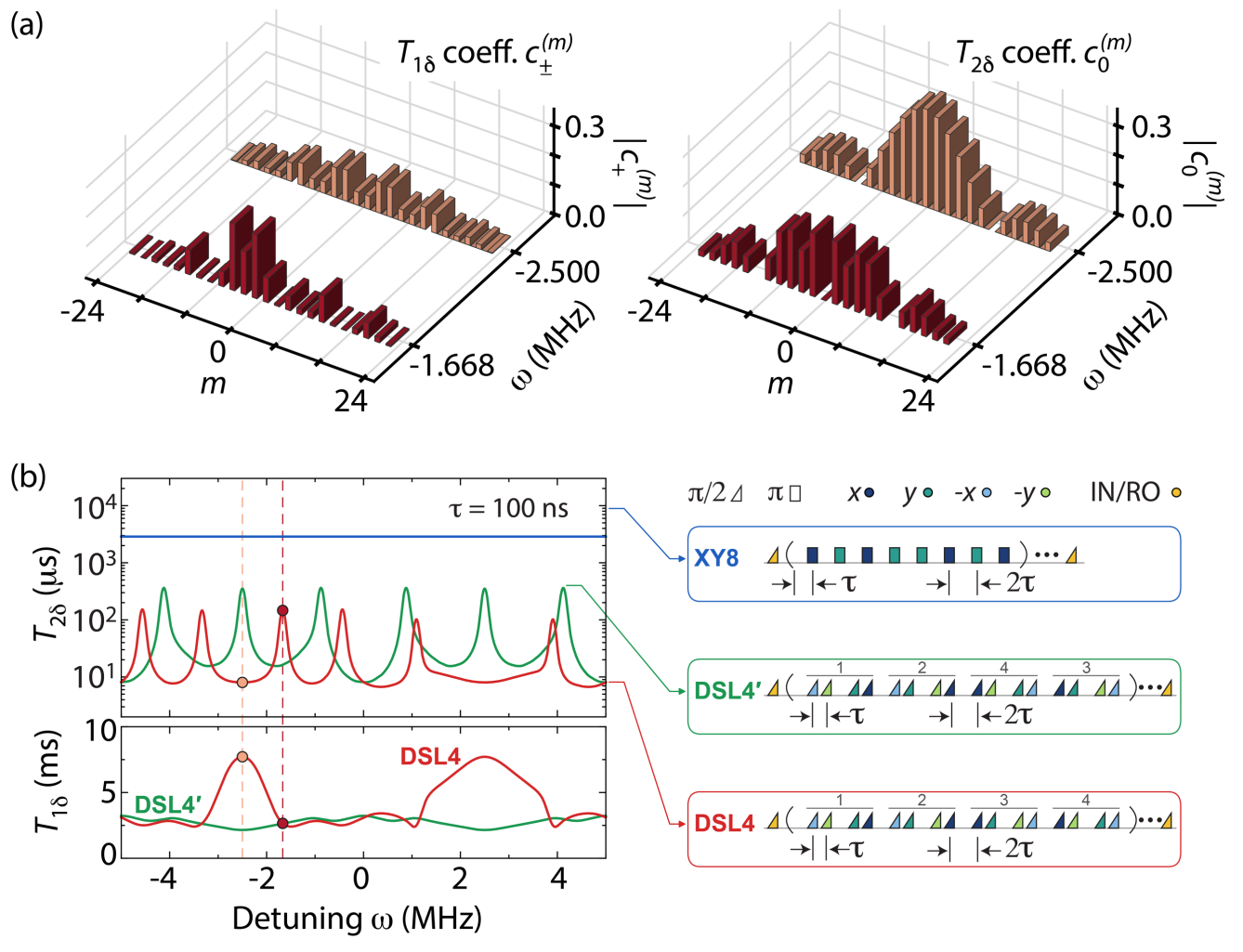

**FIG. 4: Engineering Floquet relaxation.** (a) Distribution of Fourier coefficients relevant for $T_{1\delta}$ or $T_{2\delta}$ (left and right panels, respectively). One of the chosen detunings coincide with the maximum of $T_{2\delta}$ (and consequently a minimum of $T_{1\delta}$), whereas the other corresponds to a minimum (vertical dashed lines in (b)). (b) (Top) Calculated $T_{2\delta}$ for three control sequences under the same noise spectrum: DSL4 (red), XY8 (blue), and DSL4' (green), a permutation of the experimental sequence illustrating how the sequence defines the $T_{2\delta}$ profile. (Bottom) $T_{1\delta}$ for the family of DSL4-like protocols. Throughout all calculations, we assume ideal pulses; the color notation for pulse phases and notation in (b) follows Fig. 1a.

Changing the control sequence redistributes these contributions without modifying the environment. Figure 4b shows an example simply obtained by permuting the pulse order in the original DSL-4 sequence. The resulting $T_{1\delta}$ and $T_{2\delta}$ profiles differ substantially showing how periodic driving provides a versatile means to reshape the decay channels through which a quantum system samples environmental fluctuations. More generally, systematic searches over pulse protocols could be used to suppress selected higher-harmonic $m \neq 0$ coupling coefficients at targeted detunings, potentially extending coherence protection to levels comparable to (or exceeding) those possible via known dynamical decoupling approaches (e.g., XY8 in the upper plot of Fig. 4b).

In summary, we have developed a Floquet framework for engineering relaxation and decoherence under arbitrary periodic control and demonstrated it experimentally with a single NV center. By resolving the system-environment interaction into control-dependent Floquet channels, we identify and quantitatively reproduce operating points where dephasing is strongly suppressed, and show more generally how periodic driving redistributes environmental fluctuations among distinct dephasing and population-relaxation pathways.

While our experiments are limited to NV centers, the approach can be extended to quantum dots, superconducting qubits, cold atoms and other quantum systems with dynamics describable via the spin-1/2 formalism. Besides coherence protection, this framework opens several intriguing directions. For example, the frequency dependence of the engineered relaxation channels could be exploited for magnetometry[29,30], where external fields are detected through changes in relaxation rather than coherent phase accumulation. More generally, the Floquet decomposition provides an alternative approach to quantum noise spectroscopy, potentially allowing to interrogate the environment in regimes where dynamical decoupling does not perform efficiently (e.g., in the low-frequency limit). Further, because the coefficients $c_{\alpha,\beta}^{(m)}$ provide a forward map from a control sequence to the governing interaction Hamiltonian (connecting the probe to the bath or to other spin qubits), the formalism establishes a natural basis for inverse control. In particular, extending this approach to interacting systems should enable the simultaneous engineering of coherent many-body Hamiltonians, particularly with the assistance of reinforcement-learning-based approaches[28,31].

We thank Diego Bussandri for helpful discussions. D.C., and L.B. acknowledge support from Secyt-UNC 32520230100029CB and CONICET PIP 11220200102451CO. K.X. and C.A.M. acknowledge support by the U.S. Department of Energy, Office of Science, National Quantum Information Science Research Centers, Co-design Center for Quantum Advantage (C2QA) under contract number DE-SC0012704. D.P. acknowledges support from the National Science Foundation via grant NSF-2506082. All authors also acknowledge access to the facilities and research infrastructure of the NSF CREST IDEALS, grant number NSF-2112550.

# Supplementary Material for
# "Floquet engineering of quantum relaxation"

**Kang Xu[1], Dionisio Cendoya[3,4], Nazar Delegan[5,6], F. Joseph Heremans[5,6], Olga Shenderova[7], Daniela Pagliero[1,2], Lisandro Buljubasich[3,4,†], and Carlos A. Meriles[1,2,†]**

*[1]Department of Physics, CUNY- The City College of New York, New York, NY 10031, USA.*
*[2]CUNY-Graduate Center, New York, NY 10016, USA.*
*[3]Facultad de Matemática, Astronomía, Física y Computación, Universidad Nacional de Córdoba, Córdoba, Argentina.*
*[4]CONICET, Instituto de Física Enrique Gaviola (IFEG), Córdoba, Argentina.*
*[5]Materials Science Division and Q-NEXT, Argonne National Laboratory, Lemont, IL 60439, USA.*
*[6]Pritzker School of Molecular Engineering, University of Chicago, Chicago, IL 60637, USA.*
*[7]Adamas Nanotechnologies, Inc., Raleigh, NC 27617, USA.*

[†]E-mail: lisandro.buljubasich@unc.edu.ar, cmeriles@ccny.cuny.edu

**Index**

## I. Theoretical Framework

### I.1 Spin-lattice relaxation under a periodic drive

We write the rotating-frame Hamiltonian as

$$\mathcal{H}^* = \omega(t)I_0 + \mathcal{H}^*_{\mathrm{rf}}(t) + \mathcal{H}^*_{\mathrm{int}}\,, \tag{1}$$

where $I_0 \equiv I_z$ is the operator representing the system's global angular momentum along $z$, $\omega$ is the offset relative to the system's Larmor frequency, $\mathcal{H}^*_{\mathrm{rf}}(t) = \mathbf{\Omega}_{\mathrm{rf}}(t)\cdot\mathbf{I}$ is the (periodic) control field with $\mathbf{I}$ denoting the system's global angular momentum vector operator, and we assume $\hbar = 1$. Restricting for simplicity to transition-frequency fluctuations, we take the system-bath interaction to be $\mathcal{H}^*_{\mathrm{int}} = \lambda_0 I_0$ with $\lambda_0$ denoting an operator in the Schrödinger picture, representing fluctuations of the system's transition frequency. We now introduce the control propagator $P(t)$ such that

$$i\frac{dP(t)}{dt} = \mathcal{H}^*_{\mathrm{rf}}(t)P(t)\,. \tag{2}$$

Defining $P(0) = 1$, the solution is given by $P(t) = \mathcal{T}\exp\left(-i\int_0^t dt'\mathcal{H}^*_{\mathrm{rf}}(t')\right)$, where $\mathcal{T}$ denotes the time-ordering operator. Using $\rho(t)$ to indicate the system-bath density matrix, the Liouville equation can then be recast as

$$i\frac{d\tilde{\rho}(t)}{dt} = \left[\omega\tilde{I}_z(t) + \widetilde{\mathcal{H}}_{\mathrm{int}}(t), \tilde{\rho}(t)\right]. \tag{3}$$

In the above expression, we are using a tilde to indicate a transformation to the toggling frame, namely, $\widetilde{\Theta} = P^{\dagger}(t)\Theta P(t)$ for any operator $\Theta$. Using a similar derivation, we now remove the offset dependence via the toggling frame evolution operator $Q(t) = \mathcal{T}\exp\left(-i\int_0^t dt'\,\omega\tilde{I}_z(t')\right)$ satisfying $i\frac{dQ(t)}{dt} = \omega\tilde{I}_z(t)Q(t)$. The Liouville equation now reads

$$\frac{d\tilde{\tilde{\rho}}(t)}{dt} = -i\left[\tilde{\widetilde{\mathcal{H}}}_{\mathrm{int}}(t), \tilde{\tilde{\rho}}(t)\right], \tag{4}$$

where the double tilde denotes the transformation $\tilde{\widetilde{\Theta}} = Q^{\dagger}(t)\widetilde{\Theta}Q(t)$ for an arbitrary operator $\widetilde{\Theta}$. Notice that $Q(t)$ inherits the periodicity of $P(t)$, and from Floquet's theorem, can thus be expressed as

$$Q(t) = \mathcal{F}(t)\exp(-i\mathcal{H}_{\mathrm{F}}t), \tag{5}$$

where $\mathcal{H}_{\mathrm{F}} = \mathbf{\Omega}_{\mathrm{F}}\cdot\mathbf{I}$ is the Floquet Hamiltonian, $\mathbf{\Omega}_{\mathrm{F}}$ denotes the Floquet field, and $\mathcal{F}(t)$ is the periodic Floquet propagator defined such that $\mathcal{F}(0) = \mathcal{F}(t_{\mathrm{c}}) = 1$ for a cycle time $t_{\mathrm{c}}$.

Integrating both sides of Eq. (4) and re-inserting the solution on the right hand side of the same equation, we obtain

$$\frac{d\tilde{\tilde{\rho}}(t)}{dt} = -i\left[\tilde{\widetilde{\mathcal{H}}}_{\mathrm{int}}(t), \tilde{\tilde{\rho}}(0)\right] - \int_0^t d\tau\left[\tilde{\widetilde{\mathcal{H}}}_{\mathrm{int}}(t), \left[\tilde{\widetilde{\mathcal{H}}}_{\mathrm{int}}(\tau), \tilde{\tilde{\rho}}(\tau)\right]\right]. \tag{6}$$

Without loss of generality, we can assume the starting density matrix $\rho(0) = \rho_{\mathrm{S}}(0)\rho_{\mathrm{B}}(0)$ can be expressed as a product of system- and bath-only density operators, and that $\mathrm{Tr}_{\mathrm{B}}\{\rho_{\mathrm{B}}\lambda\}$ $\langle\lambda_0(t)\rangle$ cancels out at any given time. Therefore, after taking partial trace over the bath, we rewrite Eq. (6) as

$$\frac{d\tilde{\tilde{\rho}}_{\mathrm{S}}(t)}{dt} = -\int_0^t d\tau\;\mathrm{Tr}_{\mathrm{B}}\left\{\left[\tilde{\widetilde{\mathcal{H}}}_{\mathrm{int}}(t), \left[\tilde{\widetilde{\mathcal{H}}}_{\mathrm{int}}(\tau), \tilde{\tilde{\rho}}(\tau)\right]\right]\right\}. \tag{7}$$

Here we assume that (i) the bath is stationary, so that its correlation functions depend only on time differences, and (ii) the system-bath coupling is sufficiently weak to justify the Born–Markov approximation. In the interaction picture, the bath operator evolves as $\lambda_0(t) = e^{i\mathcal{H}_{\mathrm{B}}t}\lambda_0 e^{-i\mathcal{H}_{\mathrm{B}}t}$, where $\mathcal{H}_{\mathrm{B}}$ is the bath Hamiltonian. Equation (7) then reads

$$\frac{d\tilde{\tilde{\rho}}_{\rm S}(t)}{dt} = -\int_0^\infty d\tau \; {\rm Tr_B}\left\{\left[\tilde{\tilde{I}}_0(t)\lambda_0(t), \left[\tilde{\tilde{I}}_0(t-\tau)\lambda_0(t-\tau), \tilde{\tilde{\rho}}_{\rm S}(t)\rho_{\rm B}(0)\right]\right]\right\}$$

$$= -\left\{\int_0^\infty d\tau \; \tilde{\tilde{I}}_0(t)\tilde{\tilde{I}}_0(t-\tau)\tilde{\tilde{\rho}}_{\rm S}(t)\kappa_0(\tau) - \tilde{\tilde{I}}_0(t-\tau)\tilde{\tilde{\rho}}_{\rm S}(t)\tilde{\tilde{I}}_0(t)\kappa_0(\tau) + h.c.\right\}. \quad (8)$$

where we made use of the short duration of $\tau_{\rm c}$ to extend the integral to infinity, and $h.c$ denotes Hermitian conjugate. $\kappa_0(\tau) = \langle\lambda_0(\tau)\lambda_0(0)\rangle_{\rm B}$ is the time correlator for the bath operator $\lambda_0$, which is assumed to be dependent only on the time difference $\tau$. Notice that, since $\mathcal{F}(t)$ is periodic, it is possible to recast $I_0$

$$\tilde{\tilde{I}}_0(t) = e^{i\mathcal{H}_{\rm F}t}\mathcal{F}^\dagger(t)P^\dagger(t)I_0P(t)\mathcal{F}(t)e^{-i\mathcal{H}_{\rm F}t} = \sum_m e^{im\omega_{\rm c}t}e^{i\mathcal{H}_{\rm F}t}I_0^{(m)}e^{-i\mathcal{H}_{\rm F}t}, \quad (9)$$

where $\omega_{\rm c} = 2\pi/t_c$ is the cycle frequency, and the Fourier operators in the expansion are given by

$$I_0^{(m)} = \frac{1}{t_{\rm c}}\int_0^{t_{\rm c}} dt\, e^{-im\omega_{\rm c}t}\,\mathcal{F}^\dagger(t)P^\dagger(t)I_0P(t)\mathcal{F}(t)\,. \quad (10)$$

We now resolve each Fourier operator in the Floquet basis $J_\beta$ with $\beta = \{0,+,-\}$, namely

$$I_0^{(m)} = \sum_\beta c_{0,\beta}^{(m)}J_\beta\,, \quad (11)$$

where $c_{0,\beta}^{(m)} = {\rm Tr}\left\{J_\beta^\dagger I_0^{(m)}\right\}/{\rm Tr}\left\{J_\beta^\dagger J_\beta\right\}$ quantify the projection of the physical noise operator onto the Floquet operators at harmonic $m$. Since $e^{i\mathcal{H}_{\rm F}t}J_\beta e^{-i\mathcal{H}_{\rm F}t} = e^{i\beta\Omega_{\rm F}t}J_\beta$, the interaction-picture coupling operator becomes

$$\tilde{\tilde{I}}_0(t) = \sum_{m,\beta} c_{0,\beta}^{(m)}e^{i\Omega_{0,\beta}^{(m)}t}J_\beta, \quad (12)$$

with $\Omega_{0,\beta}^{(m)} = m\omega_{\rm c} + \beta\Omega_{\rm F}$. Substitution of Eq. (12) into Eq. (8), followed by the Floquet-secular approximation (in which terms oscillating at distinct Floquet frequencies are neglected), gives the time-independent Floquet-Lindblad equation

$$\frac{d\tilde{\tilde{\rho}}_{\rm S}(t)}{dt} = -i[\mathcal{H}_{\rm LS}, \tilde{\tilde{\rho}}_{\rm S}(t)] + \sum_m\left(\gamma_0^{(m)}\mathcal{D}[J_0] + \gamma_+^{(m)}\mathcal{D}[J_+] + \gamma_-^{(m)}\mathcal{D}[J_-]\right)\tilde{\tilde{\rho}}_{\rm S}(t)\,, \quad (13)$$

where $\mathcal{D}[\Theta]\rho \equiv \Theta\rho\Theta^\dagger - \frac{1}{2}\{\Theta^\dagger\Theta,\rho\}$ for an arbitrary operator $\Theta$, and $\gamma_\beta^{(m)} = \left|c_{0,\beta}^{(m)}\right|^2\Lambda_0\left(\Omega_{0,\beta}^{(m)}\right)$ with $\Lambda_0(\Omega) = \int_{-\infty}^{\infty} d\tau\, e^{i\Omega\tau}\kappa_0(\tau)$ denoting the two-sided noise spectral density. In the above expression, $\mathcal{H}_{\rm LS}$ denotes the bath-induced Lamb-shift Hamiltonian, which renormalizes the Floquet quasienergies but does not contribute to the decay rates considered below and can thus be ignored. Equation (13) makes explicit the distinction between the two types of dissipative channels. Terms proportional to $J_0$ preserve the Floquet-state populations and produce pure dephasing, whereas $J_\pm$ induce transitions between the two Floquet states and hence population relaxation.

We can now derive an expression for $T_1$ by considering the evolution of populations. For example, for the population in the $|+\rangle$ state of the Floquet basis set $\{|+\rangle, |-\rangle\}$, Eq. (13) yields

$$\frac{d}{dt}\langle+|\tilde{\tilde{\rho}}_{\rm S}(t)|+\rangle = \Gamma_{\uparrow\delta}\langle-|\tilde{\tilde{\rho}}_{\rm S}(t)|-\rangle - \Gamma_{\downarrow\delta}\langle+|\tilde{\tilde{\rho}}_{\rm S}(t)|+\rangle \quad (14)$$

with spin-flip rates given by

$$\Gamma_{\uparrow\delta} = \sum_m \Lambda_0(+\Omega_{\mathrm{F}} + m\omega_c) \left|c_{0,+}^{(m)}\right|^2 , \tag{15}$$

$$\Gamma_{\downarrow\delta} = \sum_m \Lambda_0(-\Omega_{\mathrm{F}} + m\omega_c) \left|c_{0,-}^{(m)}\right|^2 . \tag{16}$$

Since $\frac{d}{dt}\langle -]\tilde{\tilde{\rho}}_{\mathrm{S}}(t)|-\rangle = -\frac{d}{dt}\langle +]\tilde{\tilde{\rho}}_{\mathrm{S}}(t)|+\rangle$, we conclude

$$\frac{1}{T_{1\delta}} \equiv \Gamma_{1\delta} = \Gamma_{\uparrow\delta} + \Gamma_{\downarrow\delta} , \tag{17}$$

which is precisely the population-relaxation expression used in the main text. Meanwhile, the off-diagonal density-matrix element $\langle +]\tilde{\tilde{\rho}}_{\mathrm{S}}(t)|-\rangle$ evolves as

$$\frac{d}{dt}\langle +]\tilde{\tilde{\rho}}_{\mathrm{S}}(t)|-\rangle = -\left(\frac{\Gamma_{\uparrow\delta} + \Gamma_{\downarrow\delta}}{2} + \Gamma_{\phi\delta}\right)\langle +]\tilde{\tilde{\rho}}_{\mathrm{S}}(t)|-\rangle - i\omega_{\mathrm{LS}}\langle +]\tilde{\tilde{\rho}}_{\mathrm{S}}(t)|-\rangle , \tag{18}$$

where the pure dephasing rate is given by

$$\Gamma_{\phi\delta} = \frac{1}{2}\sum_m \Lambda_0(m\omega_c) \left|c_{0,0}^{(m)}\right|^2 . \tag{19}$$

Combining Eqs. (17) and (19), the Floquet coherence time is therefore

$$\frac{1}{T_{2\delta}} \equiv \Gamma_{2\delta} = \frac{\Gamma_{1\delta}}{2} + \Gamma_{\phi\delta} . \tag{20}$$

Equations (17) and (20) qualitatively reproduce the standard expressions for longitudinal and transverse spin relaxation except that the process can now be seen as arising from contributions centered at frequencies $\Omega_{0,\beta}^{(m)}$ with $\beta = \{0, +, -\}$. Importantly, coupling to these Fourier channels is governed by coefficients of the form $\langle \pm]I_0^{(m)}|\mp\rangle$ (or $\langle \pm]I_0^{(m)}|\pm\rangle$ ) each of which arises from a one-cycle average of $\mathcal{F}^\dagger(t)P^\dagger(t)I_0P(t)(t)\mathcal{F}(t)$ modulated by a sinusoid at frequency $m\omega_{\mathrm{c}}$ (see Eq. (10)). This result is interesting in that it points to pulse protocol design and frequency detuning as valuable tools to control the way a quantum system couples to the noise produced by the environment.

Finally, the dominant low-frequency dephasing coefficient admits a useful direct relation to the Floquet quasienergy dispersion. Applying the Floquet–Hellmann–Feynman theorem to the detuning ω gives

$$\pm\frac{1}{2}\frac{\partial\Omega_{\mathrm{F}}}{\partial\omega} = \frac{1}{t_{\mathrm{c}}}\int_0^{t_{\mathrm{c}}} dt\ \langle \pm]\mathcal{F}^\dagger(t)P^\dagger(t)I_0P(t)\mathcal{F}(t)|\pm\rangle . \tag{21}$$

where $\pm\,\Omega_{\mathrm{F}}/2$ correspond to the two Floquet quasi-energies. But then, from Eq. (10), we conclude that

$$\frac{\partial\Omega_{\mathrm{F}}}{\partial\omega} = c_{0,0}^{(0)} . \tag{22}$$

Consequently, stationary points of a continuously tracked Floquet quasienergy splitting satisfy $c_{0,0}^{(0)} = 0$ and suppress the $m = 0$ dephasing channel. Higher-harmonic dephasing terms $c_{0,0}^{(m\neq 0)}$ and population-relaxation terms $c_{0,\pm}^{(m)}$, however, generally remain finite.

## II. Experimental

### II.1 Diamond sample

The measurements were performed on nitrogen–vacancy (NV) centers in a $2 \times 2 \times 0.3$ mm$^3$ delta-doped diamond sample produced at Argonne National Laboratory. An isotopically enriched $^{12}$C layer was grown by chemical vapor deposition over an Element Six electronic-grade substrate, with a delta-doped $^{15}$N layer positioned ~25 nm below the surface to define a shallow, near-surface NV ensemble. The $^{12}$C enrichment suppresses the $^{13}$C nuclear-spin bath, yielding long free-induction dephasing times ($T_2^* \approx 2.7$ μs to 4 μs); proximity to the surface sets the longitudinal relaxation time to $T_1 \approx 5$ ms. Before mounting, the sample underwent ultrasonic cleaning in an acetone, IPA, and DI-water sequence, followed by treatment with a three-acid mixture.

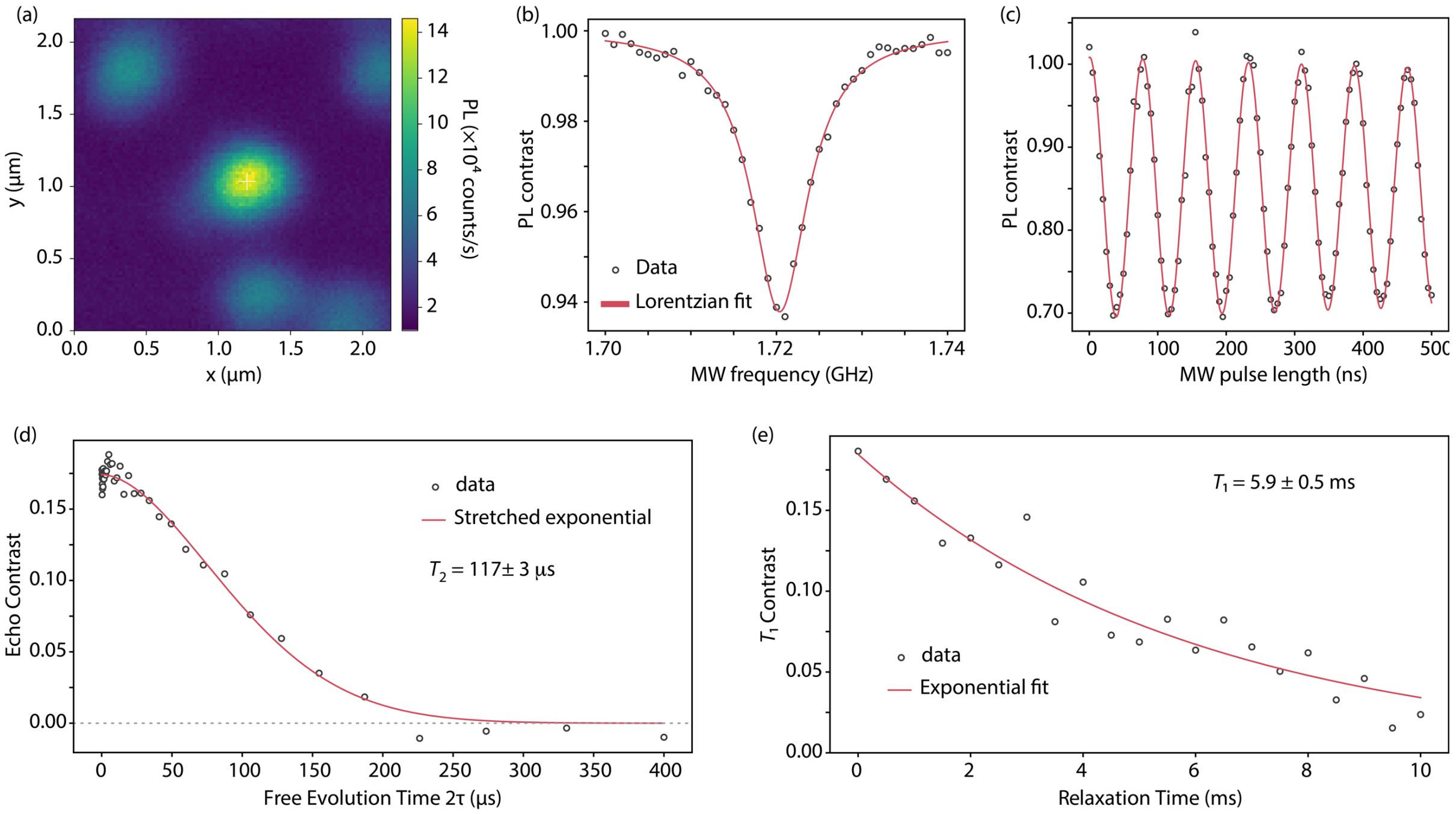

**FIG. S1: Characterization of a representative NV center.** (a) Confocal PL image of a section of the diamond with the white cross indicating the NV center investigated. (b) Optically detected magnetic resonance (ODMR) spectrum of the $|m_S = 0\rangle \leftrightarrow |m_S = -1\rangle$ transition. A Lorentzian fit (red line) yields a resonance frequency of $1720.38 \pm 0.06$ MHz corresponding to an axial magnetic field of ~41 mT. (c) Rabi oscillations driven at 1721.0MHz. A damped-cosine fit yields a Rabi frequency of $12.93 \pm 0.01$ MHz, corresponding to a π-pulse duration of ~38.7 ns. (d) Hahn-echo coherence decay as a function of the total free-evolution time 2τ. A stretched-exponential fit yields $T_2 = 117 \pm 3$ μs. (e) Longitudinal relaxation measured using differential readout with and without a π-pulse before optical detection. A single-exponential fit yields $T_1 = 5.9 \pm 0.5$ ms. In (b) and (c), the PL contrast is normalized to a reference readout; in (d) and (e), contrast denotes the differential signal between the two readout projections. All reported uncertainties correspond to one standard deviation.

## II.2 Experimental setup

Single NV centers were addressed with a home-built confocal fluorescence microscope. A 532 nm laser (~1 mW at the sample) excited the NV through a high-numerical-aperture objective, forming a diffraction-limited excitation spot. NV fluorescence was collected through the same objective, separated with a dichroic mirror, spectrally filtered by a 635 nm long-pass filter, and fiber-coupled into an avalanche photodiode (APD) for photon counting.

A static bias field was applied with a permanent magnet. Alignment along the NV symmetry axis was performed by automated motorized scanning of the magnet position while maximizing the photoluminescence (PL) contrast; the field was set so that the $|m_S = 0\rangle \leftrightarrow |m_S = -1\rangle$ transition sits at around 1721 MHz (bias field ~41 mT). As an illustration, Fig. S1 shows PL imaging, optically detected magnetic resonance (ODMR), Rabi oscillations, Hahn-echo coherence, and longitudinal relaxation measurements of a representative NV center in the set.

Periodic (Floquet) microwave driving was synthesized by a Tabor Proteus arbitrary waveform generator (AWG) operating at an 8 GHz sampling rate and delivered to the NV through a 25 μm copper wire laid against the diamond surface. The high sampling rate provides the sub-nanosecond timing resolution required to define the phase-modulated pulse train precisely. Each data point was accumulated from ~$10^3$ repetitions per shot, averaged over hundreds of experimental runs to reach adequate signal-to-noise.

### II.3 Control protocol

The pulsed control sequence — herein referred to as DSL-4 — is a phase-cycled train of resonant π/2 pulses (Rabi frequency $\Omega_{\mathrm{R}} \approx 13.2$ MHz, $t_{\pi/2} \approx 19$ ns) separated by free-evolution intervals $\tau$. The elementary building block is a permutation ("perm") cycle of four π/2 pulses spanning a period $t_{\mathrm{c}} = 6\tau$; four such perms, with the pulse phases permuted according to

$$\mathfrak{C} = (180, 270, 90, 0 \mid 180, 90, 270, 0 \mid 0, 90, 270, 180 \mid 0, 270, 90, 180)^\circ$$

form one supercycle of duration $24\tau$ (we use the phase convention: 0° = +*X*, 90° = +*Y*, 180° = −*X*, 270° =

| **Interval** | **Control $P(t)$** | **Toggling $\tilde{I}_x(t)$** | **Toggling $\tilde{I}_y(t)$** | **Toggling $\tilde{I}_z(t)$** |
|---|---|---|---|---|
| 1 | 1 | $I_x$ | $I_y$ | $I_z$ |
| 2 | $P_{\bar{x}}$ | $I_x$ | $-I_z$ | $I_y$ |
| 3 | $P_{\bar{y}}P_{\bar{x}}$ | $I_y$ | $-I_z$ | $-I_x$ |
| 4 | $P_{\bar{y}}P_{\bar{x}}$ | $I_y$ | $-I_z$ | $-I_x$ |
| 5 | $P_{\bar{x}}$ | $I_x$ | $-I_z$ | $I_y$ |
| 6 | 1 | $I_x$ | $I_y$ | $I_z$ |
| 7 | 1 | $I_x$ | $I_y$ | $I_z$ |
| 8 | $P_{\bar{x}}$ | $I_x$ | $-I_z$ | $I_y$ |
| 9 | $P_yP_{\bar{x}}$ | $-I_y$ | $-I_z$ | $I_x$ |
| 10 | $P_yP_{\bar{x}}$ | $-I_y$ | $-I_z$ | $I_x$ |
| 11 | $P_{\bar{x}}$ | $I_x$ | $-I_z$ | $I_y$ |
| 12 | 1 | $I_x$ | $I_y$ | $I_z$ |
| 13 | 1 | $I_x$ | $I_y$ | $I_z$ |
| 14 | $P_x$ | $I_x$ | $I_z$ | $-I_y$ |
| 15 | $P_yP_x$ | $I_y$ | $I_z$ | $I_x$ |
| 16 | $P_yP_x$ | $I_y$ | $I_z$ | $I_x$ |
| 17 | $P_x$ | $I_x$ | $I_z$ | $-I_y$ |
| 18 | 1 | $I_x$ | $I_y$ | $I_z$ |
| 19 | 1 | $I_x$ | $I_y$ | $I_z$ |
| 20 | $P_x$ | $I_x$ | $I_z$ | $-I_y$ |
| 21 | $P_{\bar{y}}P_x$ | $-I_y$ | $I_z$ | $-I_x$ |
| 22 | $P_{\bar{y}}P_x$ | $-I_y$ | $I_z$ | $-I_x$ |
| 23 | $P_x$ | $I_x$ | $I_z$ | $-I_y$ |
| 24 | 1 | $I_x$ | $I_y$ | $I_z$ |

**Table S1:** Explicit expressions for the control propagator $P(t)$ and toggling frame angular momentum operators $\tilde{I}_\alpha(t)$, $\alpha = x, y, z$, during subsequent time intervals $k = 1 \dots 24$ within a DSL-4 protocol. We use $P_\alpha$ to denote $\pi/2$-pulses along axis $\alpha$ (with the bar denoting the negative direction); horizontal red lines separate the different quarters.

$-Y$). This four-perm construction places the effective locking axis in the *xz* plane. Table S1 summarizes the control propagator and the corresponding toggling-frame angular momentum operators for each successive time interval of the DSL-4 supercycle, providing the explicit operator transformations used in the Floquet analysis of Section I.1.

For each experiment, the spin was initialized by a 532 nm laser pulse and a preparation π/2-pulse into the transverse plane, driven through $N$ pulses of the DSL-4 train, and read out by a final π/2-pulse that maps the transverse coherence back onto the $|m_S = 0\rangle/|m_S = -1\rangle$ populations for optical detection. The preparation and readout axes are matched to define two measurement configurations: in the *X–X* configuration the state is prepared with a $+X$ π/2-pulse and read out along $+X$ and $-X$; in the *Y–Y* configuration, it is prepared with a $+Y$ π/2-pulse and read out along $+Y$ and $-Y$.

For each configuration the measurement was repeated with the two readout phases (±*X* for *X–X*, ±Y for *Y–Y*), which project the transverse coherence onto $+Z$ and $-Z$, respectively. The two runs map the same coherence onto the $|m_S = 0\rangle$ and $|m_S = -1\rangle$ populations with opposite sign, and we report their normalized PL contrast, $\mathcal{C} = (I_+ - I_-)/(I_+ + I_-)$. This differential scheme cancels common-mode, spin-independent background (laser-power drift, readout offset, and charge-state fluctuations) and yields a background-free measure of the transverse spin projection along the prepared axis.

Two-dimensional maps were acquired by sweeping the AWG carrier frequency against the number of driving pulses $N$ (16–400 in steps of 4; total evolution time given by $N \times 1.5\tau$). The Floquet stroboscopic spectra were obtained by taking the fast Fourier transform (FFT) of the PL contrast along with $N$, sampled once per perm-cycle ($\Delta N = 4$, i.e. every $6\tau$), giving a stroboscopic Nyquist frequency $1/(2t_c)$ and a supercycle alias at $1/(4t_c)$. Measurements were taken at $\tau = 100$ ns ($t_c = 600$ ns) and $\tau = 200$ ns ($t_c = 1200$ ns).

## III. Quantitative modeling of Floquet coherence

### III.1 Experimental characterization of environmental noise

To quantitatively model the detuning dependence of the Floquet coherence time, we independently characterize the environmental noise acting on the probe NV through CPMG dynamical-decoupling measurements. We employ two complementary approaches: varying the interpulse separation $\tau$ at fixed pulse number $N$ (τ-sweeps), and varying $N$ at fixed $\tau$ (*N*-sweeps). For a given interpulse separation $\tau$, the CPMG sequence acts as a narrowband filter centered approximately at $f_{\mathrm{CPMG}} = 1/2\tau$. In the first approach, changing $\tau$ shifts the filter frequency, allowing the noise spectrum to be reconstructed from individual coherence measurements. In the second, varying $N$ changes the total evolution time $t = N\tau$ while maintaining the characteristic filter frequency, allowing the corresponding decay rate to be determined directly. We use the *N*-sweep results for quantitative modeling because they provide a more straightforward determination of the long-time decay rates, while the τ-sweeps serve as an independent consistency check.

Assuming stationary Gaussian fluctuations of the NV transition frequency, the CPMG coherence is described by $S_{\mathrm{CPMG}} = \exp(-\chi_{\mathrm{CPMG}}(t))$, where

$$\chi_{\mathrm{CPMG}}(t) = 2\pi^2 \int_0^\infty df\, \Psi(f) |Y_{\mathrm{CPMG}}(f,t)|^2 \,, \qquad (23)$$

In the above expression, $\Psi(f)$ is the one-sided power spectral density of the transition-frequency fluctuations (in Hz$^2$/Hz), related to the two-sided spectral density of Section I.1 through $\Lambda_0(2\pi f) = 2\pi^2\Psi(f)$, and $Y_{\mathrm{CPMG}}(f,t) = \int_0^t dt'\, \phi_{\mathrm{CPMG}}(t') \exp(i2\pi f t')$ is the filter function associated with the control sequence. The modulation function $\phi_{\mathrm{CPMG}}(t') = \pm 1$ changes sign at each refocusing pulse, assuming ideal instantaneous rotations.

For sufficiently long evolution times and a slowly varying spectral density across the filter bandwidth, the decay approaches an exponential form, $S_{\mathrm{CPMG}} = \exp(-\Gamma_{\mathrm{CPMG}} t)$ . In the leading-harmonic approximation, the decay rate is related to the spectral density through $\Gamma_{\mathrm{CPMG}} \approx 8\Psi(f_{\mathrm{CPMG}}) = (4/\pi^2)\Lambda_0(2\pi f_{\mathrm{CPMG}})$[1,2]. The resulting frequency-dependent rates provide an independent estimate of the dynamical noise spectrum used in the Floquet relaxation calculations. More generally, the complete filter function can be retained to account for finite-duration effects and contributions from higher harmonics.

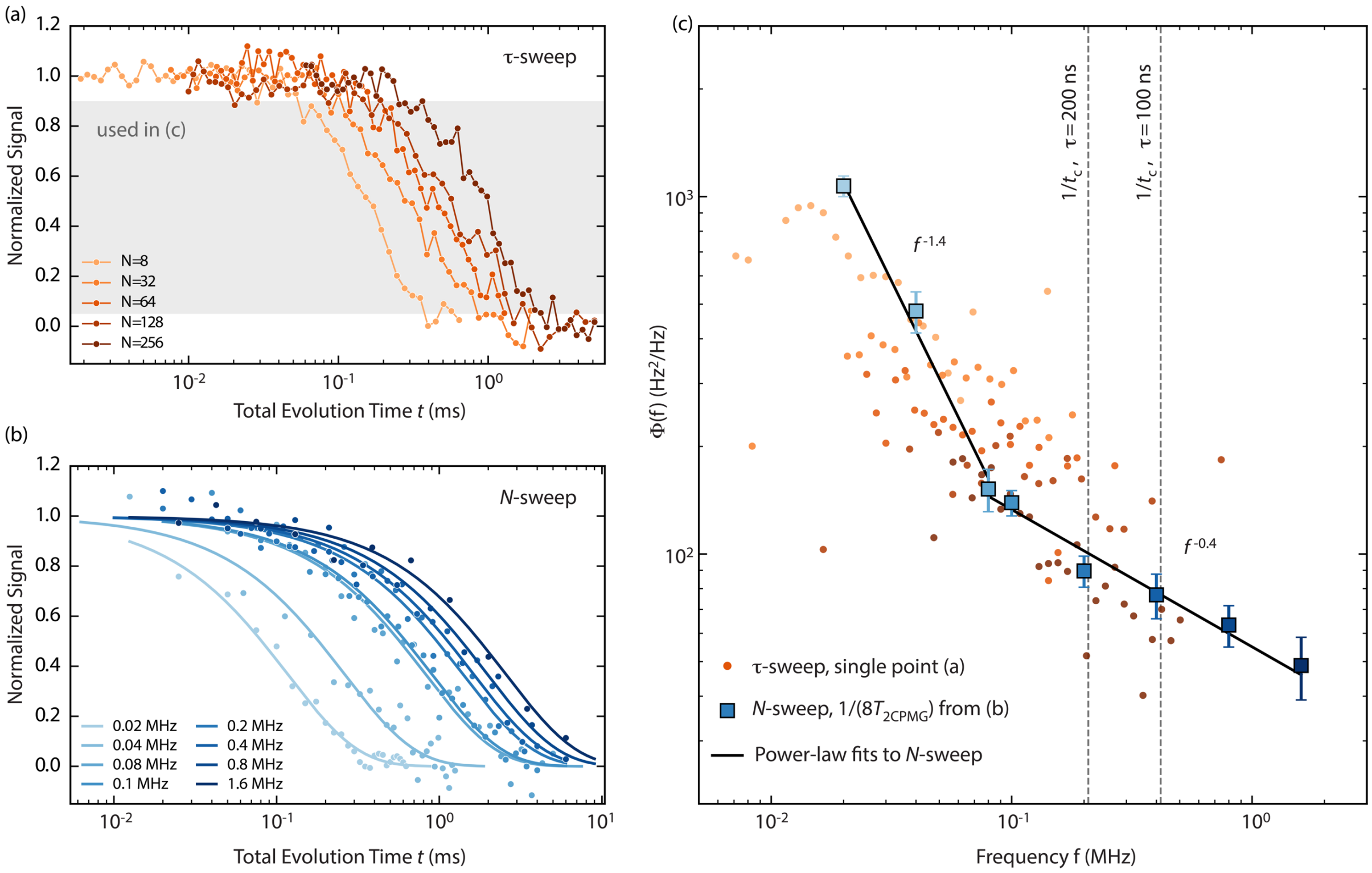

**FIG. S2: Characterization of environmental noise by dynamical decoupling.** (a) τ-sweeps: normalized CPMG coherence $S_{\mathrm{CPMG}}$ as a function of the total evolution time $t$, obtained by varying the interpulse separation at fixed numbers of refocusing pulses $N$ = 8–256 (light to dark). Points within the shaded band, $0.05 < S_{CPMG} < 0.9$, are used in (c). (b) $N$-sweeps: normalized coherence measured by varying $N$ at fixed interpulse separation $\tau$, for filter frequencies $f_{\mathrm{CPMG}}$ = 0.02–1.6 MHz (light to dark). Solid lines are exponential fits with time constant $T_{2\mathrm{CPMG}}^{(N)}$ and total time $t = N\tau$. (c) One-sided power spectral density $\Psi(f)$. Squares: $N$-sweep values, $\Psi(f_{\mathrm{CPMG}}) = 1/(8T_{2\mathrm{CPMG}}^{(N)})$, from the fits in (b), with one-standard-deviation fit errors. Circles: single-point estimates from the τ-sweep points in (a) obtained through the main lobe of the filter function (see Refs. [1] and [2]); colors as in (a) and (b). Solid lines are power-law fits to the $N$-sweep values, $\Psi \propto f^{-1.4}$ for 20–80 kHz and $\Psi \propto f^{-0.4}$ for 0.08–1.6 MHz. Dashed lines mark the DSL-4 cycle frequency $1/t_{\mathrm{c}} = 1/(24\tau)$ for $\tau = 100$ and 200 ns.

Figure S2 summarizes the CPMG measurements. In τ-sweeps (Fig. S2a), the interpulse separation is varied at fixed $N = 8$– $256$, and each data point is converted into a spectral value through the main lobe of the filter function. In $N$-sweeps (Fig. S2b), $N$ is varied at eight fixed filter frequencies, $f_{\mathrm{CPMG}} = 0.02$– $1.6$ MHz; each decay is well described by a single exponential decay with time constant $T_{2\mathrm{CPMG}}^{(N)}$. The resulting one-sided spectral densities, $\Psi(f_{\mathrm{CPMG}}) = 1/\left(8T_{2\mathrm{CPMG}}^{(N)}\right)$, decrease monotonically from $1.1 \times 10^3$ Hz$^2$/Hz at 20 kHz to 49 Hz$^2$/Hz at 1.6 MHz (Fig. S2c). They follow $\Psi \propto f^{-1.4}$ between 20 and 80 kHz and a much weaker dependence, $\Psi \propto f^{-0.4}$, between 0.08 and 1.6 MHz; this range covers the DSL-4 cycle frequencies $1/t_{\mathrm{c}} = 0.21$ MHz ($\tau = 200$ ns) and 0.42 MHz ($\tau = 100$ ns). The τ-sweep estimates agree with the $N$-sweep values to within a factor of about two over 0.02–0.4 MHz. In the calculations below, we use the $N$-sweep values, which measure the long-time decay rate directly. No correction for the finite value of $T_1$ relaxation is applied.

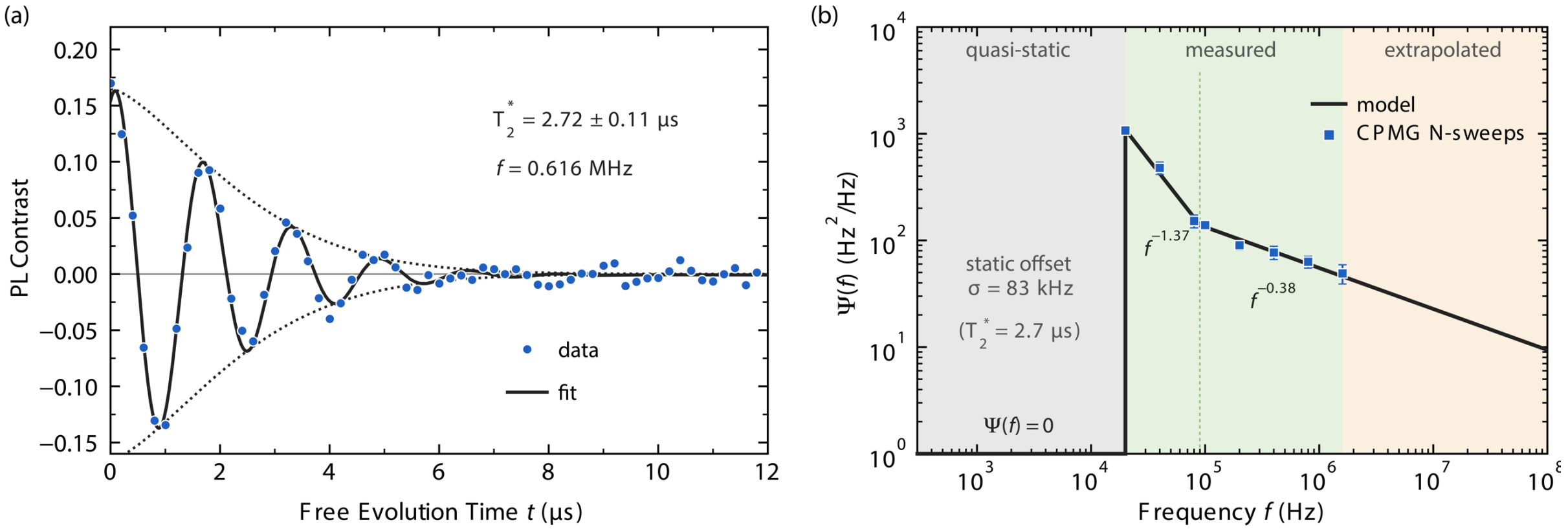

**FIG. S3: Ramsey dephasing and effective noise model for Floquet coherence calculations.** (a) Ramsey signal from the NV center used in the Floquet experiments (symbols; 46 averages) and fit to a sinusoidal damped by a stretched exponential envelope (solid line) giving $T_2^* = 2.72 \pm 0.11$ μs; the stretched exponent is $\vartheta = 1.49 \pm 0.12$. (b) Effective environmental noise model used in the Floquet coherence calculations. The dynamical contribution follows the CPMG-reconstructed spectral density within the experimentally characterized frequency interval, with phenomenological extrapolations outside this range. The additional dephasing inferred from Ramsey interferometry is represented separately by a Gaussian quasistatic detuning distribution with effective rms width $\sigma/2\pi = 83$ kHz). This approximation reproduces the measured Ramsey $1/e$ coherence time but does not attempt to reproduce its full stretched-exponential envelope.

### III.2 Ramsey dephasing and low frequency noise

The environmental noise spectrum reconstructed from CPMG measurements in Section III.1 is restricted to a finite frequency interval. To assess the contribution of fluctuations outside this interval — particularly within the low frequency range — we independently characterize the NV center using Ramsey interferometry. The measured coherence envelope follows a stretched exponential, $S_{\mathrm{R}}(t) = \exp\left(-(t/T_2^*)^{\vartheta}\right)$, with $T_2^* = 2.7 \pm 0.3$ μs and $\vartheta = 1.49 \pm 0.12$ (mean and spread of fits of the full Ramsey signal from three independent measurements; Fig. S3a). The exponent differs appreciably from the value $\vartheta = 2$ expected for Gaussian-distributed quasistatic detuning fluctuations, indicating that this limiting model does not fully describe the observed Ramsey decay. Nevertheless, for simplicity, we adopt a Gaussian-noise description in the analysis that follows, recognizing that it provides an effective approximation[3].

Within the stationary Gaussian framework invoked above, the Ramsey coherence can be expressed in terms of the one-sided spectral density $\Psi(f)$ as $S_{\mathrm{R}}(t) = \exp\left(-\chi_{\mathrm{R}}(t)\right)$, where

$$\chi_{\mathrm{R}}(t) = 2\pi^2 \int_0^\infty df\, \Psi(f) |Y_{\mathrm{R}}(f,t)|^2 \,, \tag{24}$$

with $Y_{\mathrm{R}}(f,t) = \frac{e^{i2\pi f t}-1}{i2\pi f}$. Evaluating this expression using the CPMG-reconstructed spectrum over its experimentally constrained interval — 0.02–1.6 MHz (Fig. S2c) — yields $\chi_{\mathrm{R}}(T_2^*) = 0.005$. Thus, the measured dynamical noise accounts for less than 1% of the observed Ramsey dephasing exponent at the $1/e$ coherence time. The Ramsey measurement therefore reveals an additional noise contribution not captured by the available CPMG spectrum.

Since the CPMG measurements constrain the dynamical noise spectrum over a broad frequency interval, the additional contribution responsible for Ramsey dephasing is likely dominated by fluctuations at frequencies below the experimentally accessible range. Such fluctuations can be treated as approximately quasistatic over the short Ramsey evolution time, although their behavior over the substantially longer timescales relevant to Floquet coherence remains uncertain. We therefore treat this unresolved contribution separately from the experimentally reconstructed dynamical spectrum, without assigning it a specific microscopic origin or correlation time.

This distinction is especially relevant under Floquet driving. Slow fluctuations of the NV transition frequency modify the quasienergy splitting $\Omega_F(\omega)$, and their influence on coherence is reduced near detunings satisfying $\partial_\omega \Omega_F(\omega) = 0$. The extent of this protection depends on the noise correlation times and the higher-order detuning dependence of the quasienergy splitting. The treatment of these contributions in the numerical Floquet coherence model is discussed immediately below.

### III.3 Numerical modeling of Floquet coherence

We calculate the Floquet coherence using the quasienergy spectrum and Fourier coefficients introduced in Section I.1. For each detuning $\omega$, numerical diagonalization of the single-cycle propagator yields the quasienergy splitting $\Omega_F(\omega)$, while Fourier decomposition of the transformed spin operators determines the coefficients $c_{0,\beta}^{(m)}(\omega)$. These quantities specify both the sensitivity of the dressed spin to fluctuations in its resonance frequency and the coupling strengths of the Floquet relaxation channels.

The contribution from dynamical noise is calculated using the environmental spectral density reconstructed from the CPMG measurements of Sec. III.1, converted to $\Lambda_0(\Omega) = 2\pi^2\Psi(\Omega/2\pi)$ for use in Eqs. (15), (16) and (19). The resulting transverse coherence decay rate is $\Gamma_{2\delta}^{\mathrm{dyn}} = \frac{\Gamma_{1\delta}}{2} + \Gamma_{\phi\delta}$, where $\Gamma_{1\delta}$ and $\Gamma_{\phi\delta}$ are evaluated from the Floquet transition frequencies and Fourier coefficients according to Section I.1. Here, the dynamical dephasing rate excludes the zero-harmonic contribution proportional to $\Lambda_0(0)$, which is instead accounted for through the quasistatic detuning average discussed below. No additional noise-amplitude parameter is introduced for frequencies directly constrained by the CPMG measurements.

Because the reconstructed spectrum covers only a finite frequency interval, its continuation beyond that interval must be specified separately. To account for the additional dephasing revealed by Ramsey interferometry, we adopt a Gaussian quasistatic approximation for the unresolved noise contribution. The corresponding detuning distribution is $p(\Omega) = \frac{1}{\sqrt{2\pi}\sigma}\exp\left(-\frac{\Omega^2}{2\sigma^2}\right)$. Matching the measured Ramsey $1/e$ coherence time gives an effective rms width $\sigma = \sqrt{2}/T_2^*$ where we use $T_2^* = 2.7$ μs ($\sigma/2\pi = 83$ kHz). This approximation reproduces the characteristic Ramsey decay time but not its measured stretched-exponential form (stretched exponent $\vartheta \approx 1.5$); accordingly, $\sigma$ should be regarded as an effective modeling parameter.

In the quasistatic limit, the NV transition frequency experiences a detuning offset $\Omega$ that remains constant during an individual experimental realization but varies between realizations. Importantly, this offset modifies not only the Floquet quasienergy splitting $\Omega_F$ and the relaxation rates, but also the orientation of the effective Floquet axis $\hat{\mathbf{\Omega}}_F$. We therefore calculate the coherence by averaging the full spin evolution over the normalized detuning distribution $p(\Omega)$. For a nominal detuning $\omega$, we consider an initial spin polarization along a unit vector $\hat{\mathbf{e}}_1(\omega)$ perpendicular to $\hat{\mathbf{\Omega}}_F$. In the DSL-4 protocol, the Floquet axis lies in the *xz* plane, allowing us to choose $\hat{\mathbf{e}}_1 = \hat{\boldsymbol{y}}$ for all detunings. For a particular offset $\Omega$, the initial polarization can be decomposed into components parallel and perpendicular to the shifted Floquet axis $\hat{\mathbf{\Omega}}_F(\omega + \Omega)$. These components decay at rates $\Gamma_{1\delta}$ and $\Gamma_{2\delta}^{\mathrm{dyn}}$, respectively, while the transverse component precesses at the shifted quasienergy splitting $\Omega_F(\omega + \Omega)$. The ensemble-averaged polarization is thus

$$\langle \boldsymbol{S}_F^{qs}\rangle(t,\omega) = \int_{-\infty}^{\infty} d\Omega\, p(\Omega)\Big\{a_\Omega e^{-\Gamma_{1\delta}(\omega+\Omega)t}\,\hat{\mathbf{\Omega}}_\Omega + e^{-\Gamma_{2\delta}^{dyn}(\omega+\Omega)t}\big[\cos(\Omega_F(\omega+\Omega)t)\left(\hat{\boldsymbol{e}}_1(\omega) - (a_\Omega)\hat{\mathbf{\Omega}}_\Omega\right) + \sin(\Omega_F(\omega+\Omega)t)\,\hat{\mathbf{\Omega}}_\Omega \times \hat{\boldsymbol{e}}_1(\omega)\big]\Big\} \quad (25)$$

where, for compactness, we define $\hat{\mathbf{\Omega}}_\Omega \equiv \hat{\mathbf{\Omega}}_F(\omega + \Omega)$ and $a_\Omega \equiv \hat{\boldsymbol{e}}_1(\omega)\cdot\hat{\mathbf{\Omega}}_\Omega$. The Floquet coherence envelope is then obtained from the magnitude of the polarization transverse to the nominal Floquet axis,

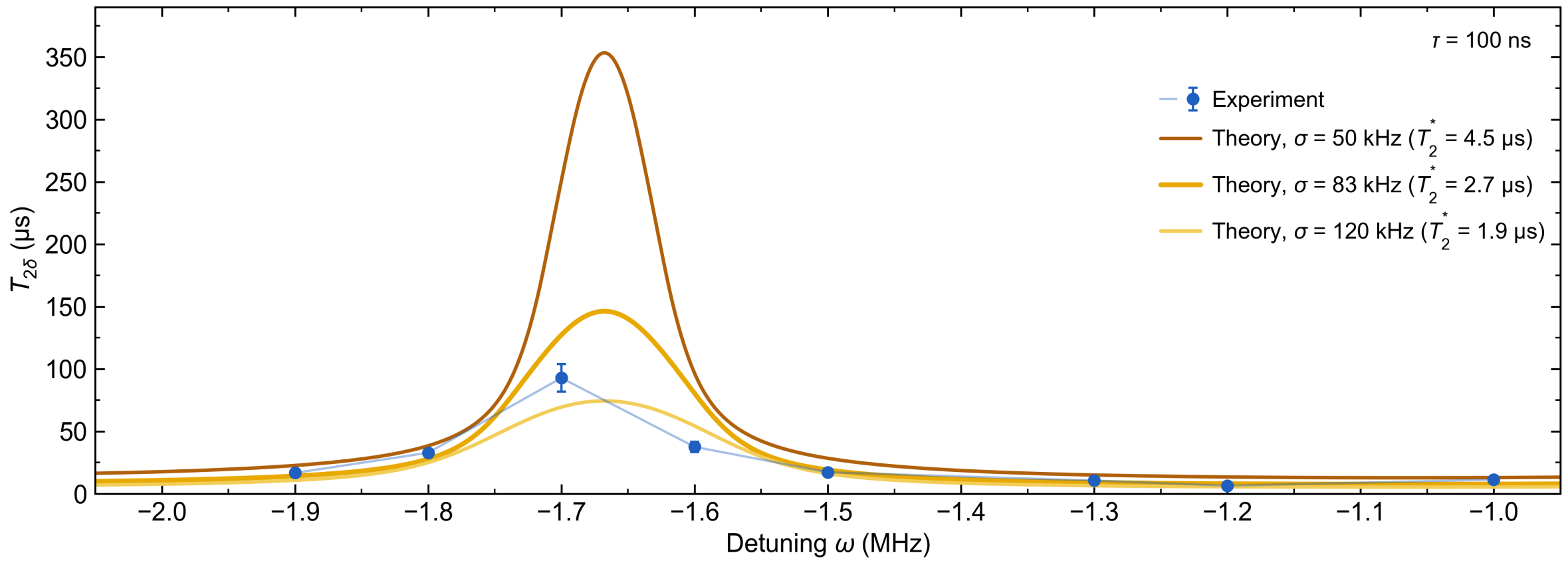

**FIG. S4: Numerical reconstruction of Floquet coherence.** (a) Experimental Floquet coherence times (symbols) as a function of microwave detuning, compared with calculated profiles (solid lines) obtained from the Floquet quasienergy spectrum, the dynamical relaxation rates inferred from CPMG noise spectroscopy, and a Gaussian quasistatic detuning distribution. The calculations are shown for three effective rms detuning widths, $\sigma/2\pi = 50, 83$ and $120$ kHz, with all other parameters held fixed. The reference value of $83$ kHz is inferred from the measured Ramsey coherence time under the Gaussian quasistatic approximation. Increasing the assumed noise amplitude reduces the predicted coherence enhancement and modifies the widths of the coherence maxima, while their approximate detuning positions remain unchanged. The comparison illustrates both the principal features captured by the simplified model and its sensitivity to the unresolved low-frequency noise contribution.

$$S_{\mathrm{F}}^{\mathrm{qs}}(t,\omega) = \left|\langle \boldsymbol{S}_{\mathbf{F}}^{\mathbf{qs}}\rangle(t,\omega) - (\langle \boldsymbol{S}_{\mathbf{F}}^{\mathbf{qs}}\rangle(t,\omega)\cdot \widehat{\boldsymbol{\Omega}}_{\boldsymbol{F}}(\omega))\,\widehat{\boldsymbol{\Omega}}_{\boldsymbol{F}}(\omega)\right| \tag{26}$$

This expression accounts for the nonlinear detuning dependence of the quasienergy splitting, relaxation rates, and Floquet axis orientation. When the axis varies negligibly across the quasistatic detuning distribution, it reduces to the simpler scalar coherence average. For the parameters considered here, including the axis variation changes the calculated $T_{2\delta}$ by at most approximately 2% for $\tau = 100$ ns and 6% for $\tau = 200$ ns.

For each nominal detuning, $T_{2\delta}$ is extracted from the calculated envelope using the same $1/e$ criterion employed in the experimental analysis. The resulting coherence enhancement reflects the combined influence of quasienergy protection and Floquet relaxation, including residual dephasing near first-order protected points where $\partial_\omega \Omega_{\mathrm{F}} = 0$. Our simplified model is intended to capture the principal features of this response, particularly the locations and overall structure of the coherence maxima, rather than provide a complete quantitative description of the environmental noise. While the peak positions are primarily determined by the Floquet Hamiltonian, their amplitudes and widths depend on the assumed quasistatic noise distribution and the extrapolation of the dynamical spectrum beyond the experimentally characterized range. A more complete treatment incorporating finite-correlation-time fluctuations is left for future work.

For the DSL-4 protocol, the Floquet quasienergy splitting and the coefficients $c_{0,\beta}^{(m)}$ are calculated numerically from the experimentally implemented pulse sequence. The resulting detuning-dependent relaxation rates are obtained by summing over the relevant Floquet harmonics, as described in Section I.1. For each nominal detuning ω, we then evaluate the coherence envelope $S_{\mathrm{F}}^{\mathrm{qs}}(t,\omega)$ introduced above using the effective Gaussian width $\sigma/2\pi \approx 83$ kHz inferred from the Ramsey measurement. The calculated Floquet coherence time is defined by $S_{\mathrm{F}}^{\mathrm{qs}}(T_{2\delta},\omega) = 1/e$. Repeating this procedure over the experimental detuning range yields a theoretical $T_{2\delta}(\omega)$ profile that can be compared directly with the measured coherence times.

The calculated response reflects two complementary mechanisms. First, suppression of the quasienergy sensitivity to detuning reduces dephasing near stationary points of $\Omega_{\mathrm{F}}(\omega)$. Second, the detuning-dependent Floquet coefficients redistribute the coupling to environmental noise among different spectral channels, modifying the dynamical relaxation rates. The resulting coherence maxima therefore emerge from the combined influence of quasienergy protection and Floquet relaxation, without introducing detuning-dependent fitting parameters. To assess the dependence of the calculated coherence enhancement on the phenomenological description of unresolved low-frequency noise, we additionally vary the effective quasistatic detuning width around its Ramsey-derived reference value, keeping the Floquet Hamiltonian and dynamical noise spectrum unchanged (Fig. S4). The calculations reproduce the location and overall shape of the experimentally observed coherence maximum, although its predicted amplitude depends appreciably on the assumed quasistatic noise width. In particular, the Ramsey-derived reference value $\sigma/2\pi = 83$ kHz yields a peak coherence time of approximately 150 μs, compared with the experimental value near 95 μs. Increasing the effective noise width reduces the predicted enhancement without significantly shifting its position (primarily determined by the Floquet control protocol). This sensitivity illustrates the precision and limitations of the quasistatic approximation.

---

[1] T. Yuge, S. Sasaki, Y. Hirayama, "Measurement of the noise spectrum using a multiple-pulse sequence", *Phys. Rev. Lett.* **107**, 170504 (2011).

[2] J. Bylander, S. Gustavsson, F. Yan, F. Yoshihara, K. Harrabi, G. Fitch, D.G. Cory, Y. Nakamura, J.-S. Tsai, and W.D. Oliver, "Noise spectroscopy through dynamical decoupling with a superconducting flux qubit", *Nat. Phys.* **7**, 565 (2011).

[3] Y. Romach, C. Müller, T. Unden, L.J. Rogers, T. Isoda, K.M. Itoh, M. Markham, A. Stacey, J. Meijer, S. Pezzagna, B. Naydenov, L.P. McGuinness, N. Bar-Gill, F. Jelezko, "Spectroscopy of surface-induced noise using shallow spins in diamond", *Phys. Rev. Lett.* **114**, 017601 (2015).